\documentclass[namedreferences]{solarphysics}
\usepackage[optionalrh]{spr-sola-addons}
\usepackage{epsfig}
\usepackage{graphicx}
\usepackage{color}
\usepackage{rotating}
\usepackage{url}
\usepackage[pdfborder={0 0 0 },urlcolor=blue]{hyperref}
\ifx \doiurl \undefined \def \doiurl#1{\href{http://dx.doi.org/#1}{\url{#1}}}\fi
\ifx \adsurl \undefined \def \adsurl#1{\href{http://adsabs.harvard.edu/abs/#1}{\url{#1}}}\fi

\newcommand{\etal}{{\it et al.}}

\begin{document}
\begin{article}

\begin{opening}

\title{A Comparison of Solar Cycle Variations in the Equatorial Rotation
 Rates of the Sun's Subsurface, Surface, Corona, and Sunspot Groups}

\author{J.\ Javaraiah}

\runningauthor{J. Javaraiah}

\runningtitle{Variation in the Solar Equatorial Rotation Rate}

 \institute{Indian Institute of Astrophysics, Bengaluru-560 034, India.\\
email: \url{jj@iiap.res.in}\\
}

\begin{abstract}
Using the {\it Solar Optical Observing Network} (SOON) sunspot-group
 data for the period 1985\,--\,2010, the 
variations in the  annual mean  equatorial-rotation rates of the sunspot groups 
are determined and  
compared  with  the known variations in the solar  equatorial-rotation rates 
determined from the following data: i) the plasma rotation rates
at 0.94R$_\odot$,  0.95R$_\odot$,\dots,1.0R$_\odot$ measured by   
 {\it Global Oscillation Network Group} (GONG) during the period 1995\,--\,2010, 
ii) the data on  the soft X-ray
 corona determined from {\it Yohkoh}/SXT
full disk images for the years 1992\,--\,2001,  iii) the data on small
bright coronal structures (SBCS) which  were
 traced in {\it Solar and Heliospheric Observatory} (SOHO)/EIT images during the period 1998\,--\,2006, 
 and iv)  
the  Mount Wilson Doppler-velocity measurements  during the period
 1986\,--\,2007.
 A large portion (up to $\approx 30^\circ$ latitude) 
of the mean differential-rotation profile of the
 sunspot groups   lies between  those of the 
internal differential-rotation rates at
 $0.94R_\odot$ and $0.98R_\odot$. The     
  variation in the yearly  mean equatorial-rotation rate of the sunspot groups
seems  to be lagging   that of the equatorial-rotation rate  determined 
from the GONG measurements by one to two years. The amplitude of the
  latter  is  very small. 
The  solar-cycle variation in the equatorial-rotation rate 
 of the solar corona   
 closely matches  that determined from the sunspot-group  data. 
 The  variation in the equatorial-rotation rate 
determined from the Mount Wilson Doppler-velocity data 
 closely resembles 
 the corresponding variation in the equatorial-rotation rate 
 determined from the sunspot-group
 data that  included  the 
  values of the abnormal angular motions 
($> |3^\circ|$ day$^{-1}$) of the
 sunspot groups. Implications of these results are pointed out.
\end{abstract}
\end{opening}

\section{Introduction}
 Studies on the solar-cycle variations in the solar differential  
rotation and meridional flow  are important for understanding
 the physical system that generates 
solar activity and the solar cycle~\cite{bab61,ub05,dg06,karak10}.
The solar differential rotation is  well studied, 
 in addition to the use of the Doppler-velocity 
measurements,  using the data on different solar magnetic tracers (mainly 
 the sunspots and the sunspot groups).  
The results derived from the sunspot and the sunspot-group data
  show variations in  
both the equatorial-rotation rate and the latitude gradient of the rotation
  on several 
time scales, including   time-scales close to 
 the 11-year  solar cycle
\cite{jg95,jk99,jbu05a,jbu05b,ju06,braj06}. 
 Recently,  \inlinecite{jj11} detected variations on  time-scales of 
  a  few days,  
 including  9 and 30\,--\,40 day quasi-periodicities,  in the 
coefficients of differential rotation determined from the Mount  Wilson
 Doppler-velocity data during Solar Cycle~22.
The $\approx$ 11-year period torsional oscillation 
detected by \inlinecite{hb80} from the Mount Wilson Doppler-velocity 
measurements 
 is confirmed by the helioseismic 
studies~($e.g.$, \opencite{howe00}; \opencite{abc08})

 \inlinecite{jk99}  studied  the variations in the  
coefficients of the differential rotation determined from the sunspot 
group data (1879\,--\,1976) and the Mount Wilson Doppler-velocity 
data (1962\,--\,1994). They found  considerable differences between   
the periodicities ($>$ two years)  in the 
rotational coefficients 
derived from the Doppler-velocity  and the sunspot-group
data.  \inlinecite{jub09} analyzed the  data on the solar surface
 equatorial-rotation rate
derived from the more  accurate Mount Wilson Doppler-velocity data 
 during the period 
1986\,--\,2007 and confirmed the short-term periodicities
 found by \inlinecite{jk99} 
  in the data  before the year 1996, but found no  statistically
significant 
variation  after  1996. That  is, there is a very large 
 difference between the temporal variations   of the equatorial-rotation 
rate during Solar Cycle~22 ($i.e.$ before 1995) and Cycle~23. 
Hence, it has been suspected that the frequent changes in the instrumentation
 of the 
 Mount Wilson spectrograph might have made the data 
before 1995  erroneous and  responsible for the variations in the 
equatorial-rotation rate derived from this data. To verify this, here we
compare  
the variations in the  
 equatorial- rotation rates derived from different rotational data and
 by using different techniques.
  
\section{Data and Analysis}
\subsection{\tt METHODOLOGY}
 The solar differential rotation can be determined
from the full-disk Doppler-velocity data  using the traditional  polynomial
expansion 
$$\omega(\phi) = A + B \sin^2 \phi + C \sin^4 \phi ,\eqno(1) $$
 and from  sunspot data by using  the first two terms of the expansion, $i.e.$
$$\omega(\phi) = A + B \sin^2 \phi ,\eqno(2) $$

\noindent where $\omega(\phi)$ is the solar sidereal angular velocity at
heliographic latitude $\phi$, the coefficient  $A$  represents
the equatorial-rotation rate, and  $B$ and $C$   
 measure the
latitudinal gradient in the rotation rate
with $B$ mainly representing  low latitudes and $C$ mainly
higher latitudes.

The sidereal rotation rate [$\omega$, in degrees day$^{-1}$] of a sunspot
 or sunspot group
 is computed as \cite{car63,bal86,gm79}: 
 $$\omega = \frac{\Delta L}{\Delta t} + 14.18,  \eqno(3) $$
or as ($e.g.$~\opencite{hgg84}; \opencite{kn90}; \opencite{gupta99}; 
\opencite{braj06}):    
$$ \omega = \frac{\Delta D_{CM}}{\Delta t}+ 0.9856 , \eqno(4) $$ 
\noindent where $\Delta L$, $\Delta D_{\mathrm CM}$, and  $\Delta t$ are the
differences between the heliographic longitudes [L], the central 
meridian distances [$D_{\mathrm CM}$], and the  observation times [$t$]
of two consecutive days 
observations of the sunspot or sunspot group, respectively, and  the values 
$14.18^\circ$ day$^{-1}$ and 0.9856 are the Carrington rigid-body rotation rate 
and the correction factor  corresponding to the Earth's rotation, respectively.
The quantity $\frac{\Delta D_{\mathrm CM}}{\Delta t}$ represents the synodic rotation rate. 
 Within the uncertainties, both these methods yield  the same result 
(the latter gives a relatively small value for the rotation rate; 
 see \opencite{kn90}). 
Javaraiah and co-authors have used  Equation~(3)  in all of  their  earlier 
 articles ($e.g.$~\opencite{jg95}, \citeyear{jg97a}, \citeyear{jg97b};
\opencite{jk99}; \opencite{jj03a}, \citeyear{jj05}; \opencite{jbu05a}, 
\citeyear{jbu05b}; \opencite{ju06}). 
   (Details on the method of  conversion of 
the synodic to sidereal rotation rate  can be found   in some of the above 
cited articles, $e.g.$ \opencite{braj06} and references therein.) 

The solar differential rotation can also
 be studied by binning the data into latitude intervals of reasonably 
small size, $i.e.$ without using the above equations (see the references 
above). This subject is  reviewed by a number of  authors 
($e.g.$ \opencite{jg00}  and 
references therein).

\subsection{\tt SUNSPOT GROUP DATA} 
 Here the Greenwich sunspot-group 
data during period May/1874\,--\,December/1976 and 
 the SOON 
 sunspot-group data during 1977\,--\,2010 are used.
 These data are taken 
from    the website  
 {\tt http://solarcience.msfc.nasa.gov/green\break wich.shtml}, and  
 consist of  $t$, $\phi$, 
 $L$,  
$D_{\mathrm CM}$, {\it etc.},   
for each sunspot group on each day of its observation. By using Equation~(3),
 the  $\omega$ values are determined  from the data corresponding to
 each pair of the  consecutive 
days' observations of the sunspot groups during the period 
1986\,--\,2009. 
Each year's  data are fitted 
   to  Equation~(2).  The latitudinal dependencies in the 
mean rotation rate of the sunspot groups over the whole 
period 1977\,--\,2011 (whole SOON data set) 
 is determined by 
averaging the daily values of the rotational velocities in two-degree  
 (and  $5^\circ$) latitude 
intervals and also by fitting the whole period daily data to 
Equation~(2). 
The data in both the Northern and the Southern hemispheres have been 
combined.
These calculations are also done by using  the Greenwich 
data during the period May 1874\,--\,December 1976.

Since the rotation rates of tracers  depend on the lifetimes/sizes/age
 of the tracers, the latitudinal dependence in the initial 
rotation rates (the first two days' heliographic positions of the sunspot
 groups are used) of the
 sunspot groups is also studied  by 
classifying the sunspot groups on the basis   
 of their lifetimes:   one\,--\,three days,  four\,--\,five days,
 six\,--\,eight days, 
and $>$ eight days.
 For this the combined Greenwich and SOON data are used.

The sunspot-group data that
 correspond to  $|D_{\mathrm CM}|>75^\circ$ 
in any day of the sunspot group lifetime are not used.
  In the case     
 of the study of the initial rotation rates of the sunspot groups,  
 in order to avoid ambiguity in the identifications of the
 first two days in the  lifetimes of the sunspot groups, the data during the 
entire lifetime of a sunspot group are eliminated even in any one day 
 $|D_{\mathrm CM}|>75^\circ$.
 Since the cutoff of  $D_{CM}$ 
is applied to both the Eastern and the Western sides, the sunspot groups that
 emerged at the visible side of the Sun and after certain number of days 
 went to the other side were not taken 
into account in the data sample of any of the aforementioned classes 
(Note: each disk passage of a recurrent 
sunspot group is treated as an independent sunspot group).  
   As in our previous articles,   
the data   correspond to 
  $> 2^\circ$ day$^{-1}$  latitudinal motions 
and $> 3^\circ$ day$^{-1}$  longitudinal motions
 ({\it i.e. abnormal} $\omega$ {\it values}) have been excluded.
 Ward (\citeyear{war65}, \citeyear{war66})
 found that this precaution  substantially reduces 
the uncertainties  in the results~(see~\opencite{jg95}).
However, the 
 abnormal  values of 
$\omega$ may have some physical significance. For example,  
the abnormal rotation rates of sunspot groups  may play some role  
in the productions of solar  
flares~\cite{hs03,s10}.
  Hence,   most of  the above calculations are done for 
both the cases, $i.e.$, both the sets of  the sunspot-group data 
 with and without 
the abnormal $\omega$ values included are used. 

\subsection{\tt MOUNT WILSON DOPPLER VELOCITY DATA}
  The daily values of
the equatorial-rotation rate $A$
 derived from the Mount Wilson Doppler measurements
 during the period 3 December 1985 to 5 March 2007 
 are available (\opencite{jub09}.
This period covers  Solar Cycle~22 and most of Cycle~23.
The data have been corrected for 
 scattered-light~(for  details see~\opencite{ul01}). 
 \inlinecite{jub09}  used this data  after removing 
   the very large spikes ($i.e.$ $> 2\sigma$,
 where $\sigma$ is the
standard deviation of the original time series) and studied 
short-term variations in the corrected data. 
The time series has data gaps that vary in size,
with a maximum gap of 49 days during Carrington-rotation numbers
 1560\,--\,1608. Therefore,  the daily data were binned to one-year 
consecutive intervals
 and the annual average values of the equatorial-rotation rate
 (average of the daily values of $A$  over each year) were determined.
In Figure~1 of \inlinecite{jub09} 
  both the original 
 and the corrected   time series
of the equatorial-rotation rate were shown. 
In the present analysis  the corrected time series is used.

\subsection{\tt DATA ON SUBSURFACE AND CORONAL ROTATION RATES}
The Sun's internal-rotation rates 
determined from the  GONG data for 147 intervals of three GONG-months
(a GONG-month is 36 days)   which began 
on 7 May 1995 and ended on 31 October 2009,  for 0.005R$_\odot$,
 0.01R$_\odot$, 0.015R$_\odot$,...,0.995R$_\odot$, 1.0R$_\odot$,  
and  for latitudes 0, 2, 4,...,88 degrees are available~(\opencite{ab10}). 
  Figure~1 shows  variations in the equatorial-rotation rate
 [$\Omega_0$, angular velocity at latitude $\phi = 0$]  
at 0.94R$_\odot$, 0.95R$_\odot$,..., 1.0R$_\odot$. As can be seen in this 
figure  the temporal patterns of the equatorial-rotation rates at the 
  different depths are  largely  similar. However, there is a
 suggestion that     
 some features are smoothed when going from interior towards 
the surface and {\it vice versa}. Specifically, the maxima in 2002 and 2008 are
 smoothed in deeper layers while the minimum in 2005
is smoothed from interior towards the surface.

\begin{figure}
\centerline{\includegraphics[width=\textwidth]{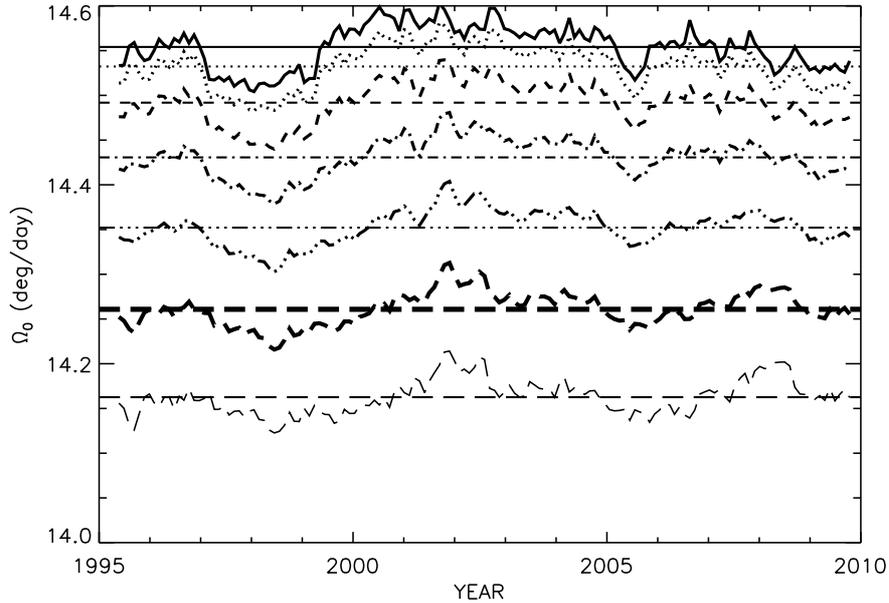}}
\caption{The solid, dotted, short-dashed, one dotted-dashed, 
three dotted-dashed, thick long-dashed, and thin long-dashed curves, represent 
 the variations  in the equatorial-rotation rate ($\Omega_0$)  
at 0.94R$_\odot$, 0.95R$_\odot$,..., 1.0R$_\odot$
 (in the order of decreasing 
mean values of $\Omega_0$),
respectively, 
determined
 from the GONG data for 147 intervals of three GONG-months (for details about this
 data  see Antia and Basu, 2010 and references therein).}
\end{figure}

 The annual mean values of the equatorial-rotation rate   of the soft X-ray corona determined from the {\it Yohkoh}/SXT solar full-disk images for the period
 1992\,--\,2001 are taken from the Table~2 of 
\inlinecite{cvi10}  and those determined from  the rotation rates of small
 bright coronal structures (SBCS) that  were
 traced in SOHO/EIT images during the period 1998\,--\,2006 are taken from 
Figure~1 of \inlinecite{jurd11}.

\section{Results}
\subsection{\tt DEPTH DEPENDENCE IN THE DIFFERENTIAL ROTATION OF 
SUNSPOT GROUPS}  

 Figure~2 shows the  latitudinal dependence in  
 the mean rotation rate of sunspot groups determined from  
 the SOON  sunspot-group data  during the period   1977\,--\,2011, 
separately from the data sets   with and without the abnormal 
$\omega$ values included. In the same figure 
 the  latitudinal dependencies in the mean
(over the whole period 1995\,--\,2009) rotation rates
of the plasma at different depths of the solar convection zone    
deduced from  the  GONG data are also shown.
As can be seen in this figure, a large portion (up to $\approx 30^\circ$
latitude) of the rotational profile that is obtained 
from the sunspot-group data 
that do not 
include the abnormal $\omega$ values lies between 
 the corresponding profiles 
at 0.94R$_{\odot}$\,--\,0.98R$_{\odot}$  determined from the  GONG data. 
(Note: The value of the correlation coefficient shown in these figures is
    almost the same for any profile of the internal rotation shown in these 
figures.)  
As can be seen in Figure~3, the similar conclusion  can also be drawn from 
the analysis of the large set of Greenwich data during
 May 1874\,--\,December 1976 (however, the helioseismic measurements 
are not exist before 1995). 
 The  latitude dependence in the rotation 
determined from the sunspot group  data that included the
 abnormal $\omega$ values is obviously highly unrealistic. Since the 
equatorial-rotation rate determined from these data seems to be somewhat 
closer to the $\Omega_0$ at 1.0R$_{\odot}$, the comparison between  its
 variation
and that of the surface is studied here. 

 There are more abnormal values in  SOON data (5.8\%)
than in Greenwich data (2.9\%). 
In the case of Greenwich data the observation time  contains the date with
 the fraction of a day. In  SOON data the fraction  is rounded to 0.5 day. 
This might some extent increased the number of abnormal $\omega$ 
values in SOON data. Figure~4 shows the 
  latitudinal dependence of the rotation determined by binning the 
sunspot-group  data that included the abnormal $\omega$ values
 into $5^\circ$ latitude intervals. Obviously, the results shown 
  in this figure the errors
 are relatively much smaller than those shown in the Figures~2(b) 
and (3(b). 
However, the latitudinal patterns of the data in $5^\circ$ 
latitude intervals  (represented by the open circles)
  are largely same as the corresponding patterns of the data in 
 $2^\circ$ latitude intervals ($cf.$, Figures~2(b) and 3(b)). 
The Greenwich data 
that included the abnormal $\omega$ values show  rigid-body rotation (this is 
 more clear in Figure~4(b) than in Figure~3(b)). 
 
\begin{figure}
\centerline{\includegraphics[width=\textwidth]{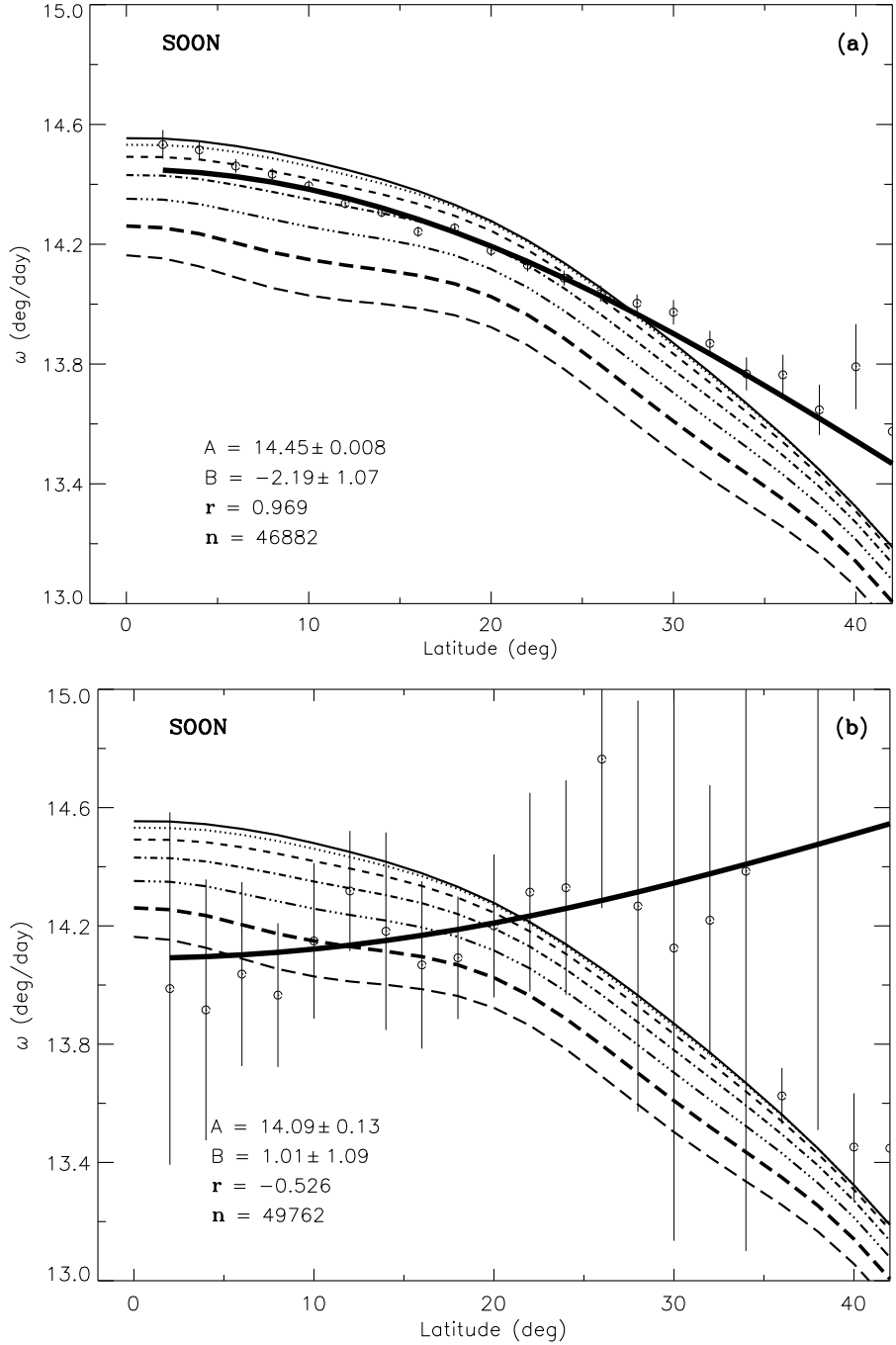}}
\caption{The latitudinal dependence of the mean rotation rates of
 sunspot groups determined by averaging the daily values of $\omega$ 
obtained from the SOON sunspot-group data during  1977\,--\,2011 over
 $2^\circ$ latitude intervals, $1^\circ - 3^\circ$,
 $3^\circ - 5^\circ$,  $5^\circ - 7^\circ$,\dots,$40^\circ - 42^\circ$
 (plotted at  $2^\circ$,  $4^\circ$,  $6^\circ$,\dots,$42^\circ$). 
The error bars represent the standard errors.
 Panels (a) and  (b) represent the data  with and without
  the abnormal values of $\omega$, respectively.
 The thick solid curve represents
the corresponding mean profile  deduced from the values (also shown) of
the coefficients
$A$ and $B$ of Equation~(2) obtained from   the total number of daily data 
 [$n$]. 
The different  type curves 
represent the  latitudinal dependencies in   the mean (over the whole period 
1995\,--\,2009)
 plasma rotation rates  deduced from  the  GONG data
 at the same  depths  as  in Figure~1.    
  $r$ represents the coefficient
 of correlation
between the sunspot group and the GONG data around 0.96R$_\odot$ in 
$2^\circ$, $4^\circ$, $6^\circ$,\dots,$36^\circ$ latitudes.}
\end{figure}

\begin{figure}
\centerline{\includegraphics[width=\textwidth]{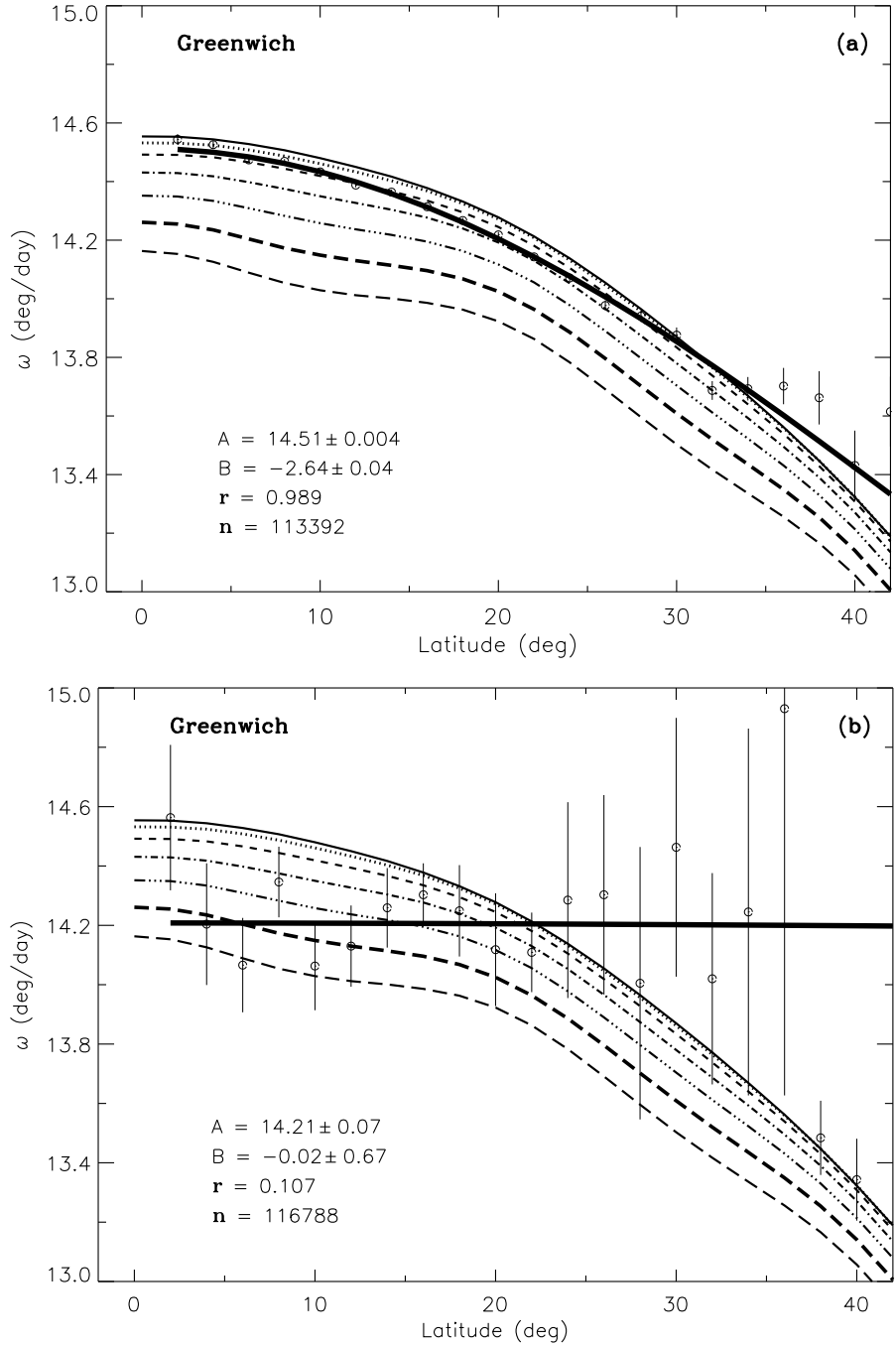}}
\caption{The latitudinal dependence of the mean rotation rates of
 sunspot groups determined by averaging the daily values of $\omega$ 
obtained from the Greenwich data during the period May 1874\,--\,December 1976
 over $2^\circ$ latitude intervals, $1^\circ - 3^\circ$,
 $3^\circ - 5^\circ$,  $5^\circ - 7^\circ$,\dots,$40^\circ - 42^\circ$
 (plotted at  $2^\circ$,  $4^\circ$,  $6^\circ$,\dots,$42^\circ$). 
The error bars represent the standard errors.
 Panels (a) and  (b) represent the data  with and without
  the abnormal values of $\omega$, respectively.
 The thick solid curve represents
the corresponding mean profile  deduced from the values (also shown) of
the coefficients
$A$ and $B$ of Equation~(2) obtained from   the total number of daily data 
 [$n$]. 
The different  type curves 
represent the  latitudinal dependencies in   the mean (over the whole period 
1995\,--\,2009)
 plasma rotation rates  deduced from  the  GONG data
 at the same  depths  as  in Figure~1.    
  $r$ represents the coefficient
 of correlation
between the sunspot group and the GONG data around 0.96R$_\odot$ in 
$2^\circ$, $4^\circ$, $6^\circ$,\dots,$36^\circ$ latitudes.}
\end{figure}

\begin{figure}
\centerline{\includegraphics[width=\textwidth]{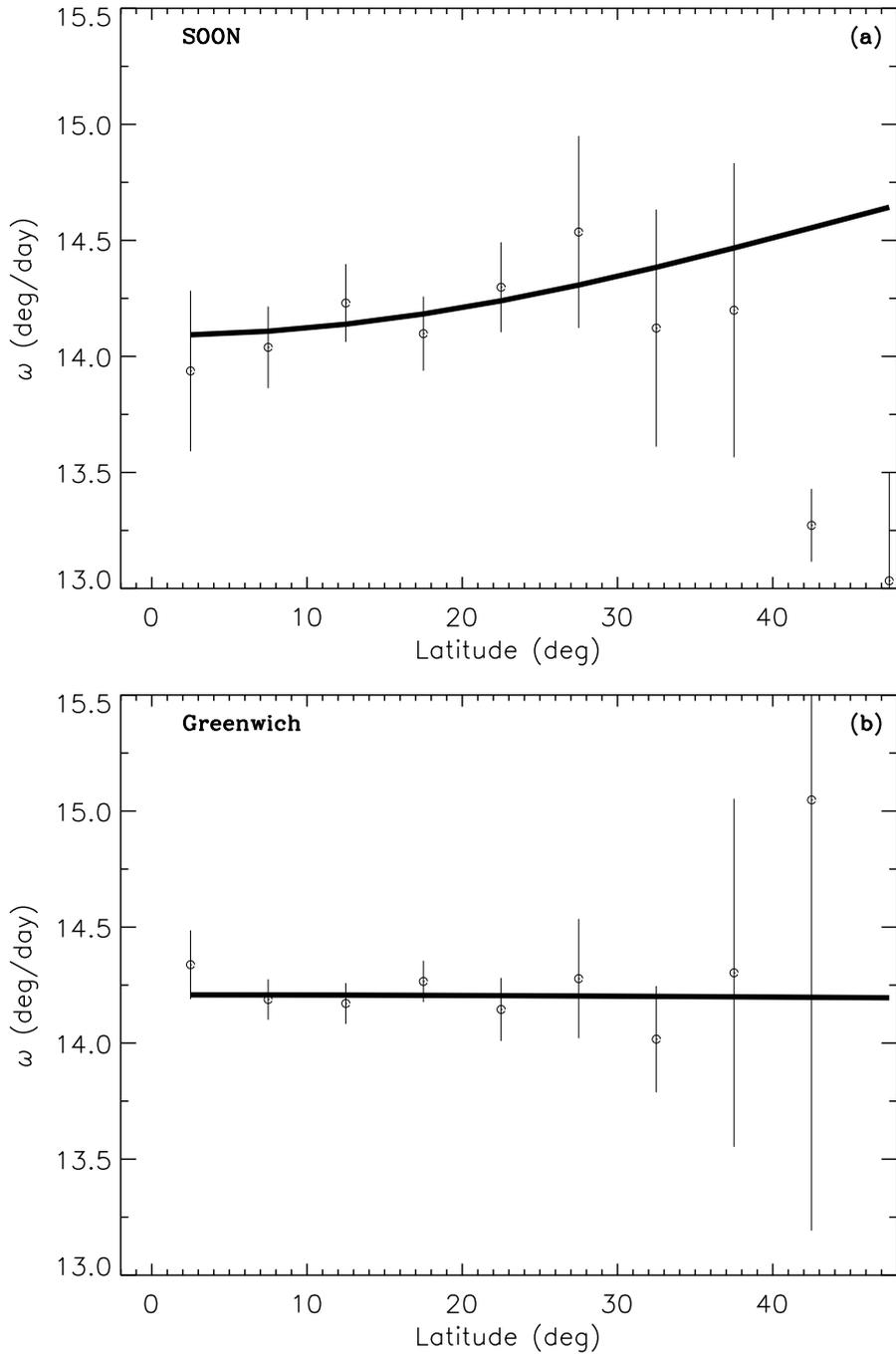}}
\caption{The latitudinal dependence of the mean rotation rates of
 sunspot groups determined by averaging the daily values of $\omega$,
 obtained from the SOON (1977\,--\,2011) and Greenwich (1874\,--\,1976) 
 sunspot-group data that included the abnormal values of $\omega$,  
 over $5^\circ$ latitude intervals, $0^\circ - 5^\circ$,
 $5^\circ - 10^\circ$,  $10^\circ -15^\circ$,\dots,$45^\circ - 50^\circ$
 (plotted at  $2.5^\circ$,  $7.5^\circ$,  $12.5^\circ$,\dots,$47.5^\circ$).
The error bars represent the standard errors.
 The thick solid curve represents
the corresponding mean profile  deduced from the values  of
the coefficients
$A$ and $B$ of Equation~(2) obtained from  the total number of 
daily data ($i.e.$ the corresponding values that are given in 
Figures~2(b) and 3(b)).}
\end{figure}

\begin{figure}
\centerline{\includegraphics[width=\textwidth]{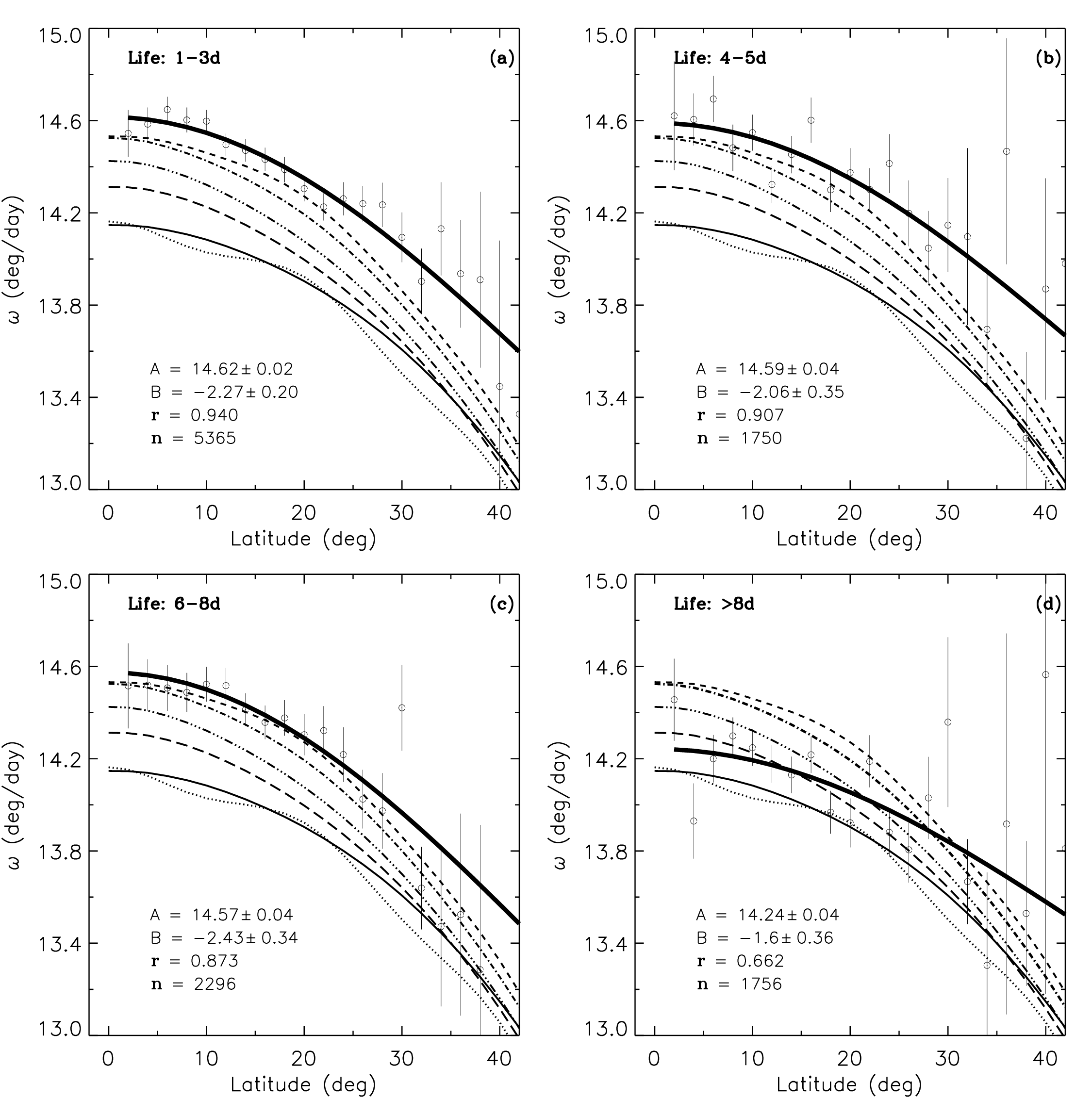}}
\caption{The latitudinal dependence of the ``mean initial rotation rates" of
 sunspot groups whose life times are in the ranges: 
(a)  one\,--\,three days, (b) four\,--\,five days, 
(c) six\,--\,eight days, and (d) $>$ eight days, determined from the combined 
 Greenwich and SOON data during the period May 1874\,--\,December 2011
  by averaging the first day values of $\omega$
 over
 $2^\circ$ latitude intervals, $1^\circ - 3^\circ$,
 $3^\circ - 5^\circ$,  $5^\circ - 7^\circ$,\dots,$40^\circ - 42^\circ$
 (plotted at  $2^\circ$,  $4^\circ$,  $6^\circ$,\dots,$42^\circ$).
The error bars represent the standard errors.
 The thick solid curve represents
the corresponding mean profile  deduced from the values (also shown) of
the coefficients
$A$ and $B$ of Equation~(2) obtained from   the total number of daily data 
 [$n$]. 
Here the internal-rotational profiles are correspond to the  different depths: 
 thin-solid, long-dashed, three-dotted-dashed, one-dotted-dashed, dashed,
 and dotted  curves
represent the  profiles  at  depths
0.75R$_\odot$, 0.80R$_\odot$, 0.85R$_\odot$, 0.90R$_\odot$,
 0.95R$_\odot$, and 1.0R$_\odot$, respectively. 
(Note: the sunspot-group data corresponding to
  $|D_{\mathrm CM}|\le 75^\circ$ are used.)}
\end{figure}

\begin{figure}
\centerline{\includegraphics[width=\textwidth]{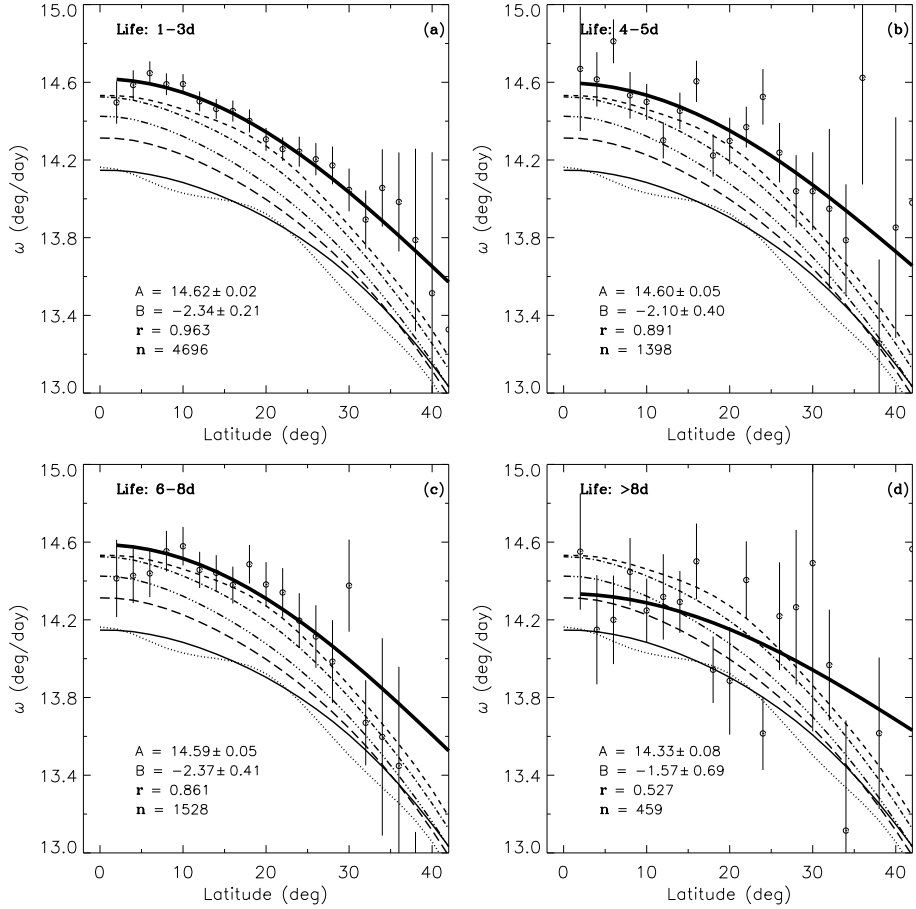}}
\caption{The  same as Figure~5, but for the sunspot-group data corresponding to 
 $|D_{CM}|\le 70^\circ$.}
\end{figure}

\begin{figure}
\centerline{\includegraphics[width=\textwidth]{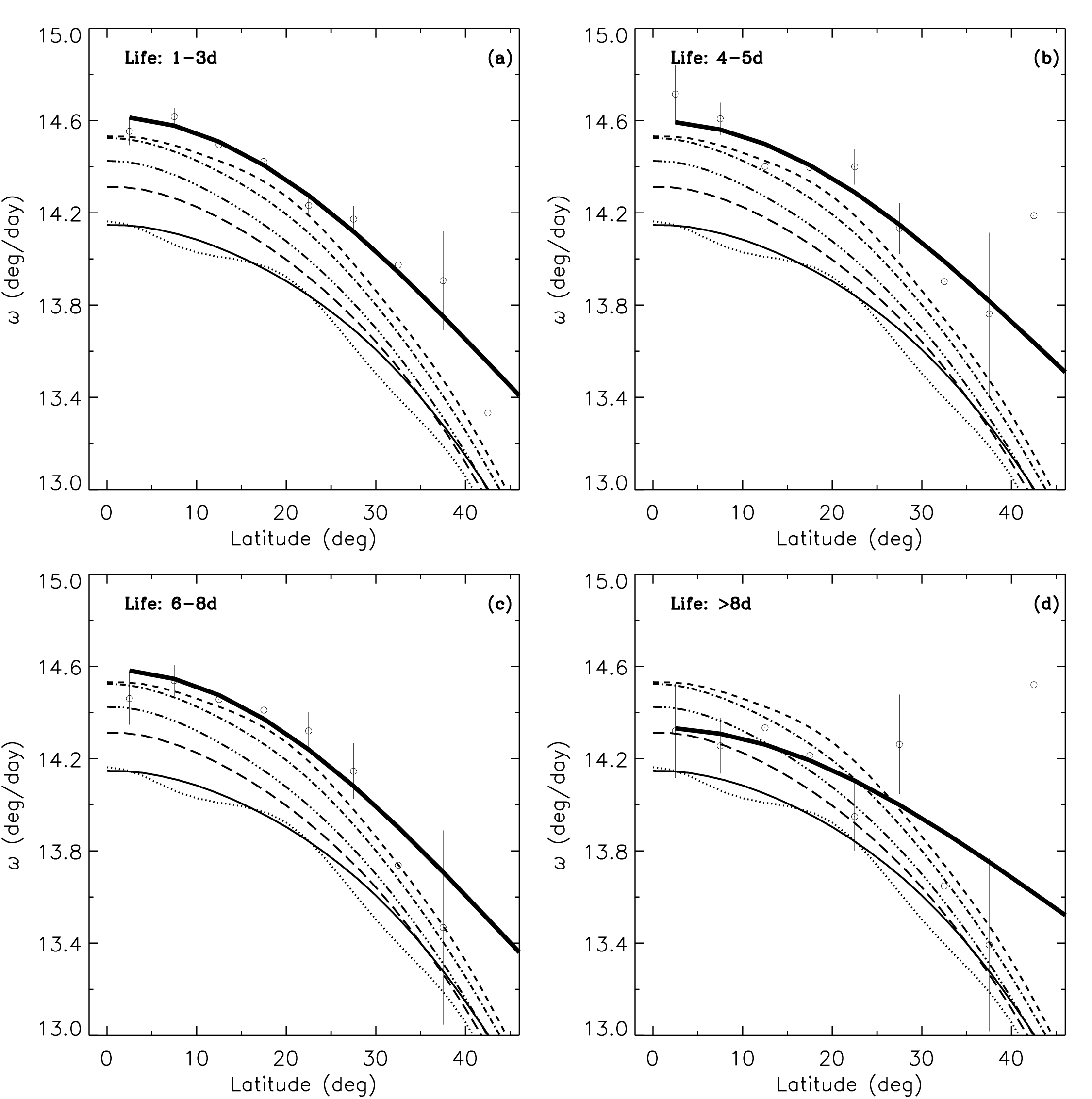}}
\caption{The latitudinal dependence of the ``mean initial rotation rates" of
 sunspot groups whose life times are in the ranges: 
(a)  one\,--\,three days, (b) four\,--\,five days, 
(c) six\,--\,eight days, and (d) $>$ eight days, determined from the combined 
 Greenwich and SOON data during the period May 1874\,--\,December 2011
  by averaging the first day values of $\omega$
 over
 $5^\circ$ latitude intervals, $0^\circ - 5^\circ$,
 $5^\circ - 10^\circ$,  $10^\circ - 15^\circ$,\dots,$45^\circ - 45^\circ$
 (plotted at  $2.5^\circ$,  $7.5^\circ$,  $12.5^\circ$,\dots,$47.5^\circ$).
The error bars represent the standard errors.
Here the internal-rotational profiles are correspond to the  different depths: 
 thin-solid, long-dashed, three-dotted-dashed, one-dotted-dashed, dashed,
 and dotted  curves
represent the  profiles  at  depths
0.75R$_\odot$, 0.80R$_\odot$, 0.85R$_\odot$, 0.90R$_\odot$,
 0.95R$_\odot$, and 1.0R$_\odot$, respectively. 
 The sunspot-group data corresponding to
  $|D_{\mathrm CM}|\le 70^\circ$ are used.
 The thick solid curve represents
the corresponding mean profile  deduced from the values  of
the coefficients
$A$ and $B$ of Equation~(2) obtained from  the total number of 
daily data ($i.e.$ the corresponding values that are given in Figure~6).}
\end{figure}

On  average  the rotation rate determined from the magnetic 
tracers is larger than that derived from the Doppler-velocity measurements.
 This difference can be interpreted  as the 
Doppler-velocity measurements  representing the surface gas motion and
 the tracers motions  representing the  motions of 
 the  deeper layers \cite{f72,nm05}, where the magnetic structures of the 
tracers are anchored. 
 Besides the concept  ``anchoring of the magnetic
 structures of sunspot groups 
to the subsurface layers", 
the faster rotation rates of sunspots, compared to the 
photospheric plasma, have been explained in several  ways:
  i) the emergent motion of the magnetic flux 
loop driven by buoyancy~($e.g.$~\opencite{mor86}; \opencite{cf89};  
\opencite{shi90}), ii) geometrical
 projection effects~\cite{gesz90}, 
 iii) the drag due to ambient 
flows~\cite{mey79,pet90},  and  iv) the interaction 
between magnetic buoyancy,   drag,  and Coriolis forces acting on the rising
 flux tubes~\cite{dsh94}.
However, in spite of the aforementioned effects, the similarity between the 
variation in the initial rotation rates of the sunspot groups with 
 their lifetimes and the radial 
variation of the internal rotation rate determined from helioseismology 
suggests that the magnetic structures of sunspot groups with successively 
longer life times (2\,--\,12 days) are initially anchored in 
successively deeper layers of the Sun's convection zone~\cite{jg97b}. 
 \inlinecite{hi02} confirmed this result   
by analyzing  a large set of the  
  Greenwich sunspot-group data and  
\inlinecite{siva03} confirmed it  
 by analyzing the sunspot-group data measured at the  
 Mount Wilson and Kodaikanal Observatories. 
 Short-lived/small sunspot 
groups predominate in a given time interval~\cite{jj12}.
 Moreover,  in the low
latitudes  near the base 
of the  convection zone to near 0.8R$_\odot$
 the internal rotation rate steeply 
increases and then it  gradually   increases  (or remains almost constant) up to
 near
 0.95R$_\odot$.  The magnetic 
structures of the sunspot groups whose lifetimes  exceed eight  days
 seem to initially anchor in the layers below 0.8R$_\odot$, 
and those of  the sunspot groups with  life times up to 8 days seem
 to initially anchor in the layers above 0.8R$_\odot$, as suggested by 
the pattern of 
the variation in the initial rotation rates of the sunspot groups with their 
lifetimes (see \opencite{jg97b}). 
In  Figure~5  the latitudinal dependencies in the mean initial rotation 
rates of the four different classes of sunspot groups, whose life times 
in the range  one\,--\,three, four\,--\,five, six\,--\,eight, and 
 $>$ eight days, are compared 
with the latitudinal dependencies in the internal rotation rates at
 depths 0.75R$_\odot$, 0.8R$_\odot$,  0.85R$_\odot$,  0.9R$_\odot$,
  0.95R$_\odot$, and 1.0R$_\odot$.   
 As can be seen in this figure, the portions up to $25^\circ$ latitude 
  in the profiles 
of the mean initial rotation rates of the shorter than  and longer than eight
 days living 
 sunspot groups 
are somewhat closer to  those of the internal rotation 
 at 0.94R$_\odot$\,--\,0.96R$_\odot$ and 
 0.8R$_\odot$, respectively. 
  These results are largely consistent with the 
results/suggestions in~\inlinecite{jg97b}.
(The profile of the very short-lived sunspot 
groups is even above the  rotation profile at 0.94R$_\odot$. The angular
 motions of super-granules may influence the rotation rates of the very 
short-lived sunspot groups. In this regard it may be worth noting that 
the magnetohydrodynamic drag force may be large on the small magnetic 
structures~(\opencite{dsh94}).  
 The results (the profiles of the mean rotation rates of  sunspot groups)
 shown in
  Figures~2 and 3 
are determined from  the combined data 
of all of the sunspot groups during all of the days in their respective 
lifetimes.
 These  results 
and also the results (the mean profiles of the 
initial rotation rates of sunspot groups) 
shown in  Figures 5(a\,--\,c) 
are similar to the results  found in
  most of  the other such studies in which   the sunspot groups were not
 sorted out according to their lifetimes~(\opencite{tnen62}; \opencite{tv88}). 
 \inlinecite{ruz04} found that 
 the initial velocity of recurrent sunspot groups is larger than the 
non-recurrent sunspot groups and suggested that 
the recurrent sunspot groups initially anchor at 0.93R$_\odot$.

 The range of the internal rotation rate that corresponds
to the  increase in depth from 
 0.94R$_\odot$\,--\,0.75R$_\odot$ is approximately the same as that 
corresponding to the 
  decrease in the depth from 0.94R$_\odot$\,--\,1.0R$_\odot$. 
In view of the result that the magnetic structures of
 the sunspot groups with successive
longer life times (2\,--\,12 days) are initially anchored in
successively deeper layers throughout the Sun's convection zone~\cite{jg97b}, 
in Figure~5  the internal rotation rates are shown for a wide range of 
depths.  In Figures~2 and 3  the mean rotation rates of the sunspot groups
are compared  with the 
 internal rotation rates at a  narrow and relatively shallower region, 
  because 
 the average rotation profile of sunspot groups 
  is mostly contributed 
from the rotation rates of  long-lived (lifetime $>$ eight days) sunspot groups
 during the last few days of their lifetimes~(\opencite{ruz04}).
The rotation rates of 
the long-lived/large sunspot groups are considerably lower during their
 initial and final 
days of their lifetimes. 

 The derived mean rotation rate of sunspot groups somewhat depends on the 
cutoff of $D_{\mathrm CM}$. It decreases when larger $D_{\mathrm CM}$ values
 are allowed 
(\opencite{ruz04} and references therein). In  all their studies 
 Javaraiah and co-authors allowed 
  $|D_{\mathrm CM}| \le 75^\circ$.  
 Ward (\citeyear{war65}, \citeyear{war66}) allowed $|D_{\mathrm CM}|$  up to 
$80^\circ$, whereas many others allowed it up to $60^\circ-70^\circ$ only.  
A larger cutoff of $D_{\mathrm CM}$ relatively  reduces the contributions of 
long-lived sunspot groups, 
particularly contributions from during their initial and final days,
 for the derived mean rotation rate. 
Hence, the derived value of the mean rotation rate of sunspot groups
 is relatively large. However, it is necessary to make sure that 
 the first days data sample should not 
contaminated with the sunspot groups arriving from other side of the Sun.     
 Thus, the calculations of the results shown 
in Figure~5 are repeated by restricting  $D_{\mathrm CM}$ up to $\pm 70^\circ$ only, 
and the  results are shown  in Figures~6 and 7.   
 These results  are largely same as those shown
 in Figure~5. 
Hence, the  conclusions/inferences drawn above from results shown in
  Figure~5 are 
 largely hold good for the results shown in Figures~6 and 7 also.    

It is widely believed that 
magnetic flux,  in the form of large  flux tubes, emerges
 to the surface presumably
  from near the base of the convection zone (where the dynamo process
is believed  to be taking place) and
 responsible for sunspots and
other solar active phenomena (see \opencite{rw92}; \opencite{gough10}).
 There are also suggestions/arguments   that the sunspots form just beneath the 
surface~(\opencite{kos00}; \opencite{kos02}) and  in different layers 
throughout the convection zone~(\opencite{brand05}). The  results above
 are largely consistent with these suggestions/arguments, but the same results 
also support the idea 
that large magnetic structures 
might be generated near the base of the solar convection zone;  
  many of the large magnetic structures may be fragmenting or branching 
into smaller structures   
while buoyant
rising through the solar convection zone, $i.e.$ small magnetic structures  
may be fragmented or branched parts of the large magnetic structures
 (\opencite{jj03b}). \inlinecite{sr05} argued that the dynamical disconnection
 of sunspots from their magnetic roots should take place  during the 
final phases of their magnetic structures' buoyant ascent towards the surface.

\subsection{\tt VARIATION IN THE EQUATORIAL ROTATION RATE}
 Figure~8  shows the   variations in the annual mean values of the
 equatorial-rotation rate 
 determined from the
Doppler-velocity data 
and the sunspot-group data that did not include the abnormal values of
$\omega$. 
 In this figure we also show
 the  variations of the
equatorial-rotation rates at  0.96R$_\odot$ {and 1.0 R$_\odot$}
determined from GONG data for each of 147 intervals of three GONG-months
 during 1995\,--\,2009,  
 the  variation in the equatorial-rotation rate  of the soft X-ray 
corona determined from the {\it Yohkoh}/SXT solar full-disk images
 for the period 1992\,--\,2001
 and from the rotation rates of SBCS
 traced in  SOHO/EIT images during the period
 1998\,--\,2006 (\opencite{jurd11}), 
 and the annual mean  sunspot number (Figure~8(b)) to study solar-cycle 
behavior of the equatorial-rotation rate.
  Since the mean value ($\approx 14.5^\circ$ day$^{-1}$) of the
 equatorial-rotation rate 
  determined from the sunspot-group data during the 
period 1985\,--\,2010 is close to 
 the mean $\Omega_0$ at 0.96R$_\odot$ ($cf.$ Figures~2 and 3), 
 the variation in the former is compared  with that of the latter, 
 besides comparing  it with the  variation at 1.0R$_\odot$
 (due to the scaling problem the  variation in $\Omega_0$ at 1.0R$_\odot$  is 
 shown in Figure~8(a)
 instead of   in Figure~8(b)).
However,  as can be seen in 
 Figure~1, the patterns of the variations of  the equatorial 
 rotation rates at different depths are  largely similar.

 As can be seen in Figure~8,   
the  mean value of the equatorial-rotation rate 
 determined from the sunspot-group data
 is substantially higher (although  it is lower  during Cycle~22 than  
 in the last 11 Solar Cycles~(\opencite{jbu05b}; \opencite{sm12})) and
the amplitude of its variation is also  about ten times higher
than that determined
from the  Doppler-velocity measurements.
The variation in the equatorial-rotation rate
 determined from the sunspot-group data 
 steeply decreased during the rising phase of the Solar Cycle~22. It   
attained  minimum  at the  maximum  of this cycle and 
 remained approximately  the same level
up to the end of the cycle, and
 then it steeply increased in the beginning of Cycle~23. The overall
pattern suggests that
there exists a quasi-11-year cycle in the equatorial-rotation rate 
(correlation coefficient $\approx\ -0.4$, 
between sunspot
number and $A$).
 At the 
 beginning of Cycle~24 the equatorial-rotation rate is  considerably 
higher than that during the last about twenty years (including the beginnings 
 of Cycles~22 and 23).
 The   pattern of the variation in the equatorial-rotation rate  is
 consistent
 with the well known result of a higher
 rotation rate during the cycle minimum than
during the maximum, found in many studies ($e.g.$ \opencite{braj06}). 
However,  it should be noted that
the minimum years of the solar cycles contain mainly small 
 sunspot groups~\cite{jj12}. 
  Small sunspot groups rotate  
faster than  large sunspot groups  
($e.g.$ \opencite{hgg84}; \opencite{jg97b}, also see Section~3.1 above).
This property of the sunspot groups most probably  has an influence on  
 the aforementioned pattern of the Solar Cycle 
variation in the surface equatorial-rotation rate
 determined from the sunspot-group data.   

 The increases of rotational velocities in
2001\,--\,2002 and 2008\,--\,2009  are most pronounced in GONG results at 1.0R$_\odot$.
 Similar increases  also exist  in the 
result  obtained by using sunspot-group data (see Figure~8(a)).
 In fact,  there is  a reasonable agreement between the pattern
 of the variations in the equatorial-rotation rates
 determined
 from the sunspot-group data and  the GONG data ($\Omega_0$).
 However, there is a considerable difference in the positions of the 
 minima of these variations,  giving  an impression of 
the internal equatorial-rotation rate
leads the equatorial-rotation rate  of the sunspot groups by one to two years
  (the reason for this is not known to us).

The annual variation in the equatorial-rotation rate
  determined from the Doppler-velocity
 data  also steeply decreased
 during the rising phase and attained minimum  at the
sunspot maximum of the Solar Cycle~22, but
 it   increased sharply from 1990 to 1993
 and suddenly dropped from 1993 to 1994. Overall
 the equatorial-rotation ration rate 
is low during Cycle~23,  but there is an indication of the
 variation in the equatorial-rotation rate is  
in phase with number of sunspots in the interval 1998\,--\,2004 (however,   
 the value of  $A$ 
 suddenly increased 
around 2005 and than decreased). 
This behavior is opposite to that in the previous cycle. 
The overall pattern of the annual variation 
indicates the existence of a  five to six-year periodicity in the equatorial-rotation rate
 during Cycle~22.

\begin{figure}
\centerline{\includegraphics[width=\textwidth]{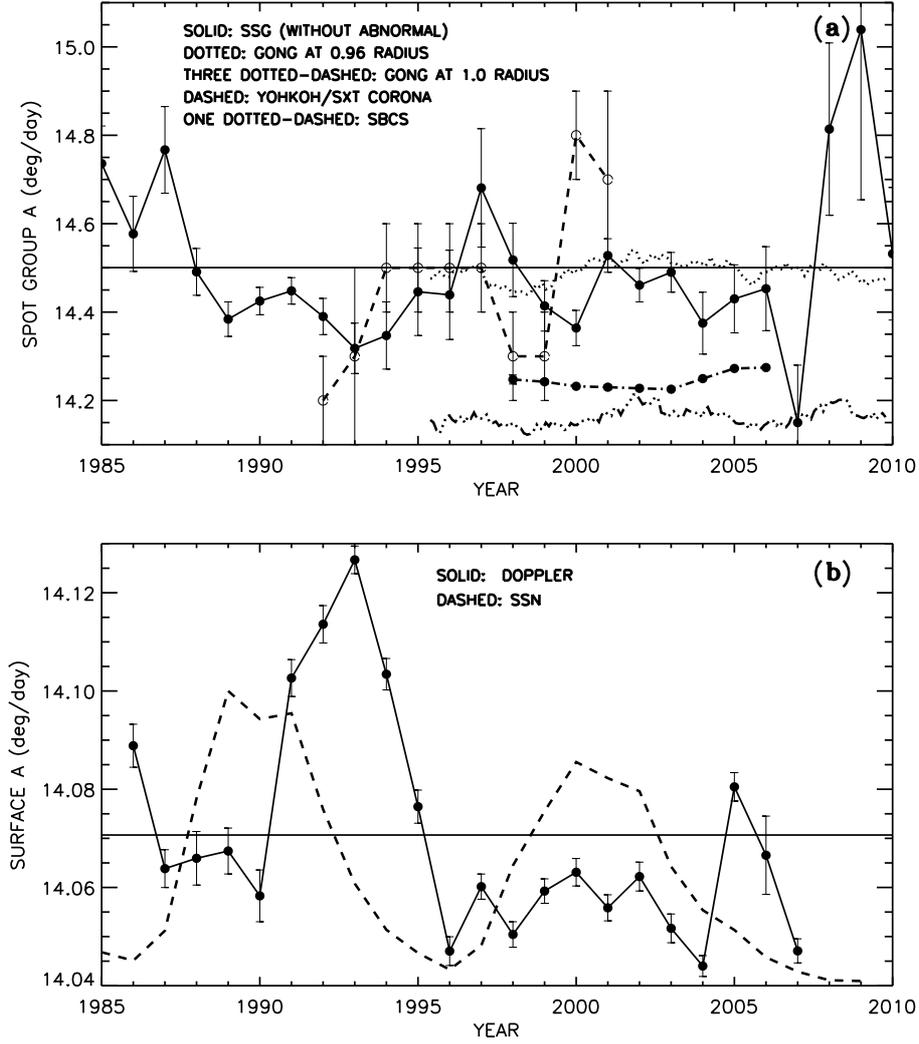}}
\caption{(a) Filled circle-solid curve: the variation in the annual
 equatorial-rotation rate $A$ determined
 from the sunspot group (SSG) data
 after  excluding the
 abnormal values of  $\omega$ which are calculated by using Equation~(1). 
 Open circle-dashed curve: 
the variations in the equatorial-rotation rate  of the soft X-ray
 corona determined from {\it Yohkoh}/SXT
full-disk images for the years 1992\,--\,2001 by Chandra, Vats, 
and Iyer (2010). Filled circle one dotted-dashed curve:  the variation in annual mean 
equatorial-rotation rate  
determined from   the data of SBCS which were
 traced in in SOHO/EIT images during the period 1998\,--\,2006 (Jurdana-\v{S}epi\'c \etal, 2011).
In each of these cases the error-bars represent the values of
 1$\sigma$  obtained from 
the linear least-square fits of the data (in the case of SBCS  $\sigma$ has very low values).
 The  dotted and three dotted-dashed curves are the  variations in the 
equatorial ration rates  at  $0.96 R_\odot$ and $1.0 R_\odot$,
determined from GONG data for each of 147 intervals of
 three GONG-months   during 1995\,--\,2009
(Antia and Basu, 2010).
The solid  horizontal line is drawn at
the  mean values of
 $14.5^\circ \pm 0.2^\circ$ day$^{-1}$,
 determined from the yearly  values of the 
 sunspot-group data.
(b) The solid curve represents the variation in the annual mean 
 $A$ determined  from
 the corrected
 Mount Wilson Doppler-velocity data  (Javaraiah $et\ al.$ 2009).
 The error bars represent one  standard error ($\sigma$ has very high values).
The   dashed curve represents the
variation in the annual mean international sunspot number (SSN), 
which is normalized
 to the scale of $A$.
The solid  horizontal line is drawn at
the  mean value of
 $A$, $14.07^\circ \pm 0.02^\circ$ day$^{-1}$,
 determined from the yearly values. 
Note: in the case of Mount Wilson Doppler-velocity data
 in  2007 the data are available only up to March.}
\end{figure}

In Figure~9 the 
variations in  the equatorial-rotation rate determined 
 from the
 Doppler-velocity data and  the sunspot-group data that included 
the abnormal values of $\omega$ ($i.e.$, $|\omega| > 3^\circ$ day$^{-1}$)  
 calculated by using Equation~(3) are compared. 
As already mentioned above,
 the inclusion of the abnormal values of $\omega$ 
 increases   the uncertainties of the derived
  coefficients $A$ and $B$,  in spite of  the 
 size of the data increasing by  
4 \%\,--\,5 \% (the highest is 8.4 \% around 2009). 
Hence, with the inclusion of abnormal $\omega$ values the errors [$\sigma$] of 
individual yearly averages are so large that they are not 
 significantly (statistically) different from each other so that the variations cannot 
be examined using them.  
Therefore,  the 
values of $A$ determined by
 binning the  data into 
 three-year moving time intervals (MTIs) are used. 
 In this
figure 
the variation in $A$ determined from the data that does not include
 the abnormal values and binned  into three-year MTIs is also shown, in order 
to get an idea about how much difference exists  between the yearly 
($c.f.$ Figure~8(a))  and 
 three-year MTIs data. 
In the case of the equatorial-rotation rate
 determined from the Doppler-velocity data,  
the three-year smoothed time series obtained from
  the annual values ($c.f.$ Figure~8(b)) is shown. 
 As can be seen in Figure~9, 
 the patterns of the variations in the equatorial-rotation rates derived
 from the  sunspot-group data and the Doppler-velocity data  closely resemble 
 one another (correlation coefficient $\approx$ 0.5). In fact, during Cycle~23  
the shapes of the curves
 are almost the same (the Doppler-velocity measurements are relatively more 
accurate after 1995).
  This result indicates that the temporal variation in the equatorial-rotation
 rate  determined from the Doppler-velocity  measurements 
is of  solar origin, which 
 has been doubtful so far for the reason mentioned in Section~1.     
However, this result/conclusion  is 
only suggestive rather than compelling, because  
the uncertainties 
in $A$ determined from the sunspot-group data 
 that included the abnormal angular motions of the sunspot groups are 
large (in the figure the error bar represent one standard error   level;  
 the corresponding  $\sigma$ is very high).
 In addition, the abnormal angular motions of the sunspot groups may
 not represent the angular motions of the surface plasma. These 
motions  may be the abnormal  proper motions of the sunspot groups,
which took place   at the locations of the eruptive
 solar phenomena such as flares. The Doppler-velocity measurements may be
 also contaminated with the contributions from  such  motions of plasma 
(Note:  the spikes in the daily Mt. Wilson 
velocity data are already removed. Further 
trimming  will create  more data gaps  and make the data more unrealistic).

\begin{figure}
\centerline{\includegraphics[width=\textwidth]{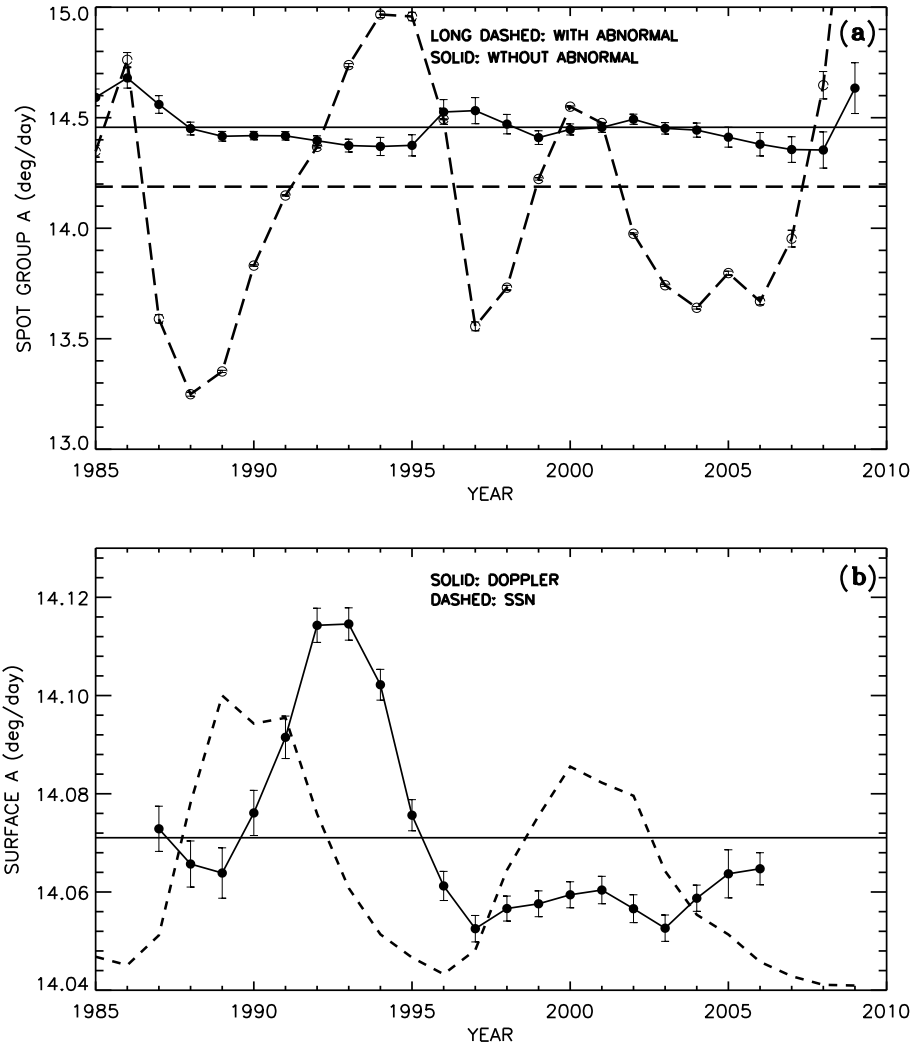}}
\caption{(a) The open circle long-dashed and the filled circle-solid curves  represent
 the variations in  $A$     derived from the sunspot
 group (SSG) data--that included  
with and   without abnormal $\omega$ values  ($> 3^\circ$ day$^{-1}$), 
respectively--in three-year MTIs successively shifted by
one year.
In the former case the error-bars 
represent the standard error (because of high $\sigma$ value) and 
in  the letter case 
 they represent the values of  1$\sigma$. 
The solid and long-dashed horizontal lines are drawn
 at the corresponding  mean values of  
 $A$, $14.46^\circ \pm 0.09^\circ$ day$^{-1}$ and $14.19^\circ \pm 0.61^\circ$ day$^{-1}$, 
respectively. (b) The filled circle-solid curve represents 
the 3-year smoothed  $A$
derived from the annual mean values of $A$ (shown in Figure~8(b))
determined from the  Mount Wilson Doppler-velocity data (error bars represent
 one standard error). 
The solid horizontal line is  drawn
 at the corresponding  mean value of  
 $A$, $14.07^\circ$ day$^{-1}$.
The   dashed curve represents the
variation in the annual mean international sunspot number (SSN),
which is normalized
 to the scale of $A$.
Note: in the case of Mount Wilson Doppler-velocity data
 in  2007 the data are available only up to March.}
\end{figure}

 On the other hand, the mean values  $14.09^\circ \pm 0.13^\circ$ and 
$14.21^\circ  \pm 0.07^\circ$ day$^{-1}$ (see Figures~2 and 3) 
of  the equatorial-rotation rates   
determined from the  SOON and Greenwich daily sunspot-group data,
 respectively,
 which   included 
the abnormal $\omega$ values 
are significantly (more than 2$\sigma$ level) lower than the
corresponding values   $14.45^\circ \pm 0.008^\circ$ and  
$14.51^\circ \pm 0.004^\circ$ 
 determined from the  data that did not include the abnormal values.
 Overall it seems that  mean values  determined from the SOON data 
that included 
the abnormal $\omega$ values   is   
somewhat closer to the mean value  ($14.07^\circ \pm 0.02^\circ$ day$^{-1}$) of 
the equatorial-rotation rate determined from 
the Doppler-velocity data. 
 In Figure~10 (the histogram of the distribution of the abnormal angular 
velocity values) there is a suggestion that  the peaks of the open 
circle-dashed curve  in 
Figure~9 
 included the majority of the  
abnormal velocity values (in the range $10^\circ - 20^\circ$) 
 that are somewhat consistent with the rotation rate of the Sun.
 Therefore,  
  the similarity in the variations of the Doppler-velocity 
data and the sunspot-group data that included the abnormal motions 
    suggest that the former may be  largely of solar origin  rather than
 spurious. 
 Nevertheless, 
 the highly unrealistic  latitudinal gradient of the 
rotation (see Figures~2 and 3) that is obtained from the sunspot-group data 
that included 
the abnormal $\omega$ values  may have a large influence on the corresponding 
 equatorial-rotation rate, making the latter  unreliable. 
 For this reason,   and because of the 
large range and the large 
 uncertainties in annual  mean values 
  (including  all errors of measurements both random 
and systematic),
  the reality/reliability of  
variation in  the sunspot-group data that included the abnormal motions 
is doubtful. 
 Hence,    
the similarity in the solar-cycle variations
 in the Doppler-velocity data and the sunspot-group data that included the
 abnormal motions  may rather suggest that variation in the 
Doppler-velocity data (before 1995) could be contributed by   the
 large uncertainties in this data.

\begin{figure}
\centerline{\includegraphics[width=\textwidth]{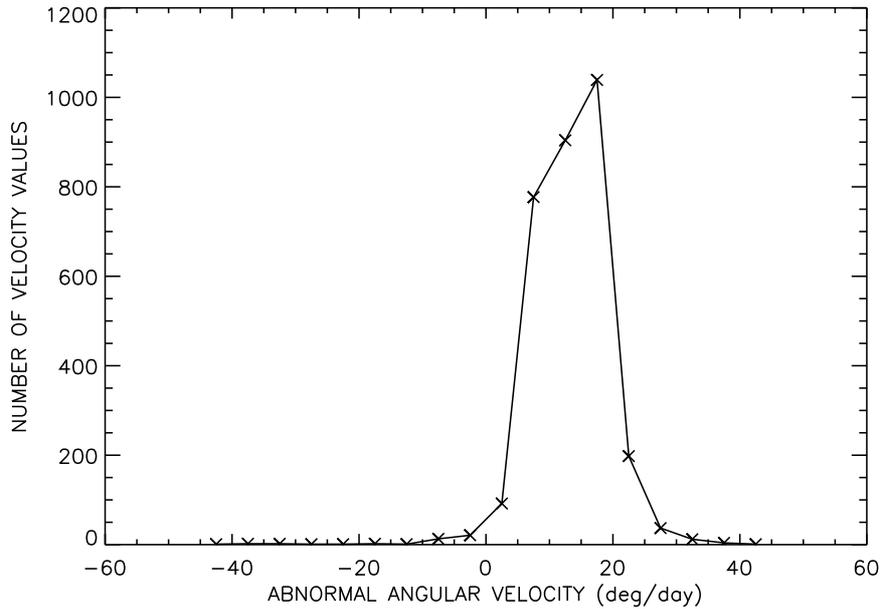}}
\caption{Distribution of the abnormal angular velocity 
values determined from the SOON sunspot-group data during the period 
 1 January 1977 to 31 January 2012.}
\end{figure}

As can be seen in Figure~8,   the equatorial-rotation rate
 determined from the Yohkoh/SXT full-disk
images for the period 1992\,--\,2001 by~\inlinecite{cvi10} 
is closely matches that determined from the sunspot group  data
  for all of the years except for the year 2000, where  it 
 has a large value.  Close to   this year
 there are humps in the variation of $A$ determined from the
  sunspot group and the Doppler-velocity data also, although they are small.   
Thus, the coronal rotation  is strongly related to the angular motions
of the  surface magnetic features.
 (There seems to be a better 
agreement between the phases of $\Omega_0$ and $A$ of the corona, particularly 
the agreement is large in minima.)
However, the pattern of $A$ determined from the SBCS seems to be
 considerably different from those of other data sets, particularly 
in the period 2001\,--\,2003.  In fact, it   seems  
anticorrelated with  that of  the GONG data.  The reason for this 
is not known.  
(It should be noted that the aforementioned 
similarity between the 
variation in the equatorial-rotation rate
  of sunspot groups/corona and the variation in the internal
 equatorial-rotation rate is only in shape. The amplitude  of the variation
 in the internal equatorial-rotation rate is relatively very small.)

\section{Conclusions}
The following conclusions can be drawn from our analysis: 
\begin{itemize}
\item[] i) A large portion (up to $\approx 30^\circ$ latitude) 
of the mean differential-rotation profile of the
 sunspot groups   lies between  those of the 
internal differential-rotation rates at
 $0.94R_\odot$ and $0.98R_\odot$. The portions  
up to $25^\circ$ latitude in the 
 mean profiles of the initial rotation rates of the up to eight days
 and longer than  
 eight days 
living sunspot groups are close to the those of the internal rotation 
near 0.96R$_\odot$ and  0.8R$_\odot$, respectively. 
\item[] ii) At the end of Cycle~23 and the
 beginning of Cycle~24 the value of the equatorial-rotation rate determined
from the sunspot-group data
is  considerably higher than that of at the beginning of Cycle~23,  Overall
 the pattern of variation in the equatorial-rotation rate during 
Cycles~22 and 23 resembles 
 the pattern of the known solar-cycle variation in the equatorial-rotation rate
 determined from 
sunspot data.  
\item[] iii) There is   a reasonable agreement between the pattern
 of the variation in the equatorial-rotation rate  
 determined from the sunspot-group data and that determined  
 from the GONG data [$\Omega_0$].
There is also an impression that the variation in the internal 
equatorial-rotation rate  
leads  the equatorial-rotation rate of  sunspot groups by one to two  years 
 (the reason for this is not known to us). 
\item[] iv) The pattern of the known solar-cycle variation in the equatorial-rotation rate
 of the solar corona   determined from the {\it Yohkoh}/SXT full-disk
images  and SBCS  closely matches with that determined from the sunspot-group
  data (except during the period 2001\,--\,2003).
  This indicates that
 the coronal rotation  is strongly related to the rotational motion
of the surface magnetic features.
\item[] v) The  variation in the equatorial-rotation rate 
determined from the Mount Wilson Doppler-velocity data substantially differs
 from the corresponding variation in the equatorial-rotation rate 
determined from the sunspot group
 data that did not include the  values of the abnormal angular motions 
($> |3^\circ|$ day$^{-1}$) of the
 sunspot groups, whereas it closely resembles the corresponding variation 
   determined from the sunspot-group data that included  the values of the
 abnormal angular motions. 
 \item[] vi) Conclusions~v) above 
  may   suggest that  the solar-cycle variation in the surface 
equatorial-rotation rate  
determined from the  Doppler-velocity measurements (before 1995) is 
 caused by the inconsistency and uncertainties in the data 
(but it needs further studies to find a definite answer).
\end{itemize}

\noindent{Acknowledgments}\ \ 
{The author thanks the anonymous referee for the detailed 
comments and useful suggestions, 
  and H.M. Antia for providing the values
 of the internal-rotation rates that he has determined from GONG data.
The author also thanks  Luca Bertello for helpful comments and suggestions, 
  B.A. Varghese for his help in making figures,   
 and the organizers of the LWS/SDO-3/SOHO26/GONG-2011 workshop for kindly
 providing a partial financial support to attend the workshop.
 This work utilizes data obtained by the Global Oscillation
Network Group (GONG) Program, managed by the
National Solar Observatory (NSO), 
  which is
operated by AURA, Inc. under a cooperative agreement with the
National Science Foundation.
The data were acquired by instruments operated by the Big
Bear Solar Observatory, High Altitude Observatory,
Learmonth Solar Observatory, Udaipur Solar Observatory,
Instituto de Astrof\'isico de Canarias, and Cerro Tololo
Interamerican Observatory.}

{}

\end{article} 
\end{document}